\documentclass[preprint, pre]{revtex4-1}

\usepackage{graphicx}

\usepackage{amssymb,amsfonts,amsmath}

\def \>{\rangle} 
\def \<{\langle} 
 
\def\be{\begin{equation}} 
\def\ee{\end{equation}} 
\def\longrightharpoonup{\relbar\joinrel\rightharpoonup}
\def\longleftharpoondown{\leftharpoondown\joinrel\relbar}

\def\longrightleftharpoons{
  \mathop{
    \vcenter{
      \hbox{
      \ooalign{
        \raise1pt\hbox{$\longrightharpoonup\joinrel$}\crcr
	  \lower1pt\hbox{$\longleftharpoondown\joinrel$}
	  }
      }
    }
  }
}

\newcommand \bea {\begin{eqnarray}} 
\newcommand \eea {\end{eqnarray}}

\begin{document}

\title{Dynamical quorum-sensing and synchronization of nonlinear oscillators coupled through an external medium}

\author{David J. Schwab}
\affiliation{Dept. of Molecular Biology and Lewis-Sigler Institute, Princeton University, Princeton, NJ 08854}
\author{Ania Baetica}
\affiliation{Dept. of Mathematics, Princeton University, Princeton, NJ 08854}
\author{Pankaj Mehta}
\affiliation{Dept. of Physics, Boston University, Boston, MA 02215}

\begin{abstract}
Many biological and physical systems exhibit population-density dependent transitions to synchronized oscillations in a process often termed ``dynamical quorum sensing". Synchronization frequently arises through chemical communication via signaling molecules distributed through an external media. We study a simple theoretical model for dynamical quorum sensing: a heterogenous population of limit-cycle oscillators  diffusively coupled through a common media. We show that this model exhibits a rich phase diagram with four qualitatively distinct mechanisms fueling population-dependent transitions to global oscillations, including a new type of transition we term ``dynamic death''. We derive a single pair of analytic equations that allows us to calculate all phase boundaries as a function of population density and show that the model reproduces many of the qualitative features of recent experiments of BZ catalytic particles as well as synthetically engineered bacteria. \end{abstract}

\maketitle

Unicellular organisms often undertake complex collective behaviors in response to environmental and population cues. A beautiful example of this phenomenon is the population-density dependent transition to synchronized oscillations observed in communicating cell populations recently termed  Òdynamical quorum sensingÓ \cite{mehta2010approaching}.  Density-dependent synchronization has been observed in a wide variety of biological systems including suspensions of yeast in nutrient solutions \cite{de2007dynamical}, starving cellular colonies of the social amoeba {\it Dictyostelium} \cite{gregor2010onset}, and synthetically engineered bacteria \cite{danino2010synchronized}. Such transitions have also been observed in experimental studies of electrochemical oscillators and Belousov-Zhabotinsky (BZ) catalytic particles \cite{taylor2009dynamical,  tinsley2010dynamical}.

Previous theoretical work has shown that oscillators coupled through quorum sensing can display synchronized oscillations \cite{mcmillen2002synchronizing, garcia2004modeling,Russo}. Recently, a dynamic quorum sensing transition was found \cite{de2007dynamical} in a simple model of coupled identical limit-cycle oscillators introduced to study synchronization in yeast populations. Additionally, numerical studies of BZ catalytic particles indicate that heterogeneity in oscillator populations leads to interesting new phenomenon \cite{taylor2009dynamical, matthews1991dynamics, tinsley2010dynamical}. The study of population-density dependent synchronization in heterogeneous populations of oscillators is still in its infancy, in stark contrast to oscillators with direct coupling where many analytic results are available \cite{strogatz1991stability, matthews1991dynamics, strogatz2000kuramoto}. 
 
 In this paper, we consider a large population of limit-cycle oscillators with a distribution of natural frequencies, coupled diffusively through a common external media. Our work generalizes earlier models \cite{de2007dynamical} and exhibits extremely rich dynamics as the coupling strength, population density, and frequency distribution are varied. We derive several analytic results and find that model exhibits a rich phase diagram. In particular, there exist four qualitatively different mechanisms leading to synchronization as a function of coupling strength and population density:  (1) a Kuramoto-like incoherence to coherence transition, (2) an amplitude death transition due to oscillator heterogeneity, (3) a loss of both global and individual oscillations due degradation in the external media, and (4) a new type of transition we term ``dynamic death'' where the external media dynamics are not fast enough to support global oscillations. We show that the model reproduces many qualitative features observed in recent experiments on heterogeneous populations of BZ catalytic particles \cite{taylor2009dynamical} as well as synthetically engineered bacteria \cite{danino2010synchronized}.
 
To illustrate this diverse set of phenomena, we introduce a simple model of  $N\gg 1$ coupled limit-cycle oscillators where the amplitude and phase of individual oscillators are represented by  a complex number $z_j$, $(j=1\ldots N)$, with natural frequency $\omega_j$. The oscillators are diffusively coupled to an external media, represented by a complex number $Z$, through a coupling $D$. Furthermore, when chemicals leave the oscillators and enter the medium, they are diluted by a factor $\alpha=V_{int}/V_{ext} \ll 1$, which is the ratio of the volume of the entire system to the that of an individual oscillator. The external media is degraded at a rate $J$. The dynamics of the system are captured by the equation
\bea
\frac{dz_j}{dt}&=& (\lambda_0+i \omega_j - |z_j|^2)z_j -D(z_j-Z) \nonumber \\
\frac{dZ}{dt} &=& \alpha D \sum_j( z_j -Z) -J Z  \nonumber
\eea
where the $\omega_j$ are drawn from a distribution $h(\omega)$ which we assume to be an even function about a mean frequency $\omega_0$.  By introducing a dimensionless density, $\rho= \alpha N$, and shifting to a frame rotating with frequency $\omega_0$, we can rewrite the equations above as
  \bea
 \frac{dz_j}{dt}&=& (\lambda_0+i \omega_j - |z_j|^2)z_j -D(z_j-Z)  \nonumber  \\
\frac{dZ}{dt} &=& \frac{\rho D}{N} \sum_j( z_j -Z) -(J+i\omega_0) Z,
 \label{MainEq}
  \eea
 where the frequencies $\omega_j$ are now drawn from an even distribution $g(\omega)$ with mean zero.
 
 To build intuition for the system, it is helpful to consider the special case of a homogenous population where $g(\omega)$ is a delta function and all $\omega_j=0$ in (\ref{MainEq}). This model was used previously \cite{de2007dynamical} to model dynamical quorum sensing in yeast suspensions. For homogenous populations, the equations for all the $z_j$ are identical and there are two possible behaviors. The individual oscillators are quiescent with $Z=z_j=0$ or there are synchronized oscillations. We can compute the stability of the oscillator death state by linearizing the system around $z_j=Z=0$ and computing the eigenvalues, $\mu$, of the corresponding  linearized system.  
In this calculation, since all of the oscillators are identical, the dynamics are completely specified by two differential equations, one for the mean-field parameter $z=\frac{1}{N}\sum_j z_j =z_j$ and one for $Z$. Oscillator death is stable when Re$(\mu)<0$ for all eigenvalues. The corresponding requirement that the trace be negative implies $D > \lambda_0$ in the oscillator death phase. Furthermore, the characteristic equation for the eigenvalues takes the form $(\mu+A)(\mu+B+i\omega_0)=\rho D^2$, with 
 $B=D\rho+J$ and $A=D-\lambda_0$. To find the phase boundary, we plug in $\mu=a+ib$ and separate the characteristic  equation into real and imaginary parts,
 \begin{eqnarray}
a+B=\frac{\rho D^2(a+A)}{(a+A)^2+b^2} \label{HomoEq} \\
b+\omega_0=\frac{-\rho D^2 b}{(a+A)^2+b^2}.
\label{HomoEq}
\end{eqnarray}
This allows us to solve for $b$ as a function of $a$, $b(a)=-\frac{\omega_0(a+A)}{2a+B+A}$ and plug this into (\ref{HomoEq}). The resulting equation can be analyzed graphically plotting the left and right sides of (\ref{HomoEq}) as a function of $a$  \cite{nextpub}.  Since the characteristic equation is quadratic, there are two solutions, a solution with negative $a$ which guarantees the stability of the external medium, and a second solution which can change sign depending on parameters. Examining the aforementioned plot, it is clear that if the left-hand side of (\ref{HomoEq}) is greater than the right-hand side at $a=0$, then the second solution must also be negative. Thus, the amplitude death phase is stable when
 \be
 \frac{(D\rho+J)(D-\lambda_0)}{\rho D^2}  \ge \frac{(\rho D +J + D-\lambda_0)^2}{(\rho D +J + D -\lambda_0)^2+\omega_0^2},
\label{HomoSEQ}
 \ee
 where we have rewritten $A$ and $B$ in terms of the original parameters of the model.
 
  \begin{figure}[t]
\includegraphics[width=7cm]{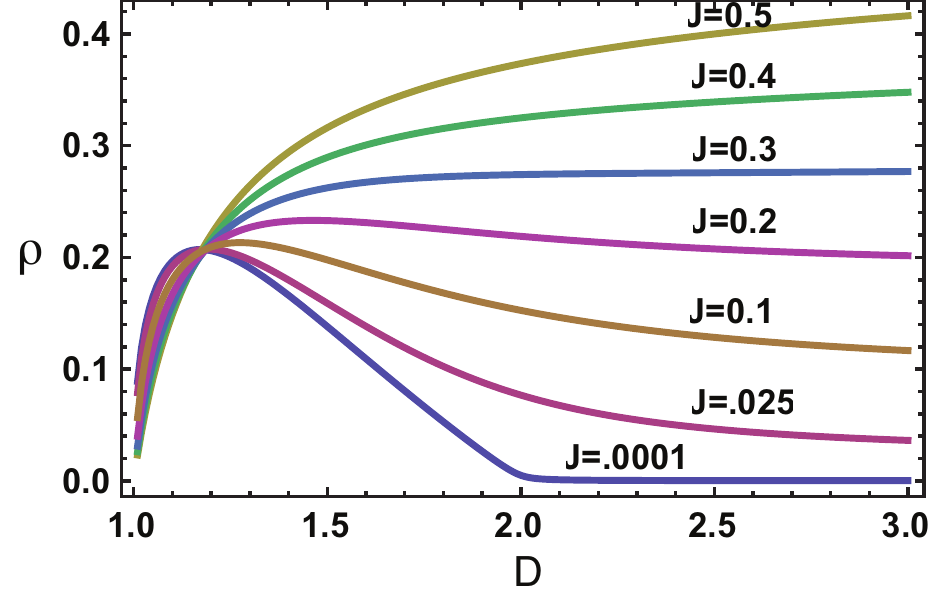}
\includegraphics[width=8cm]{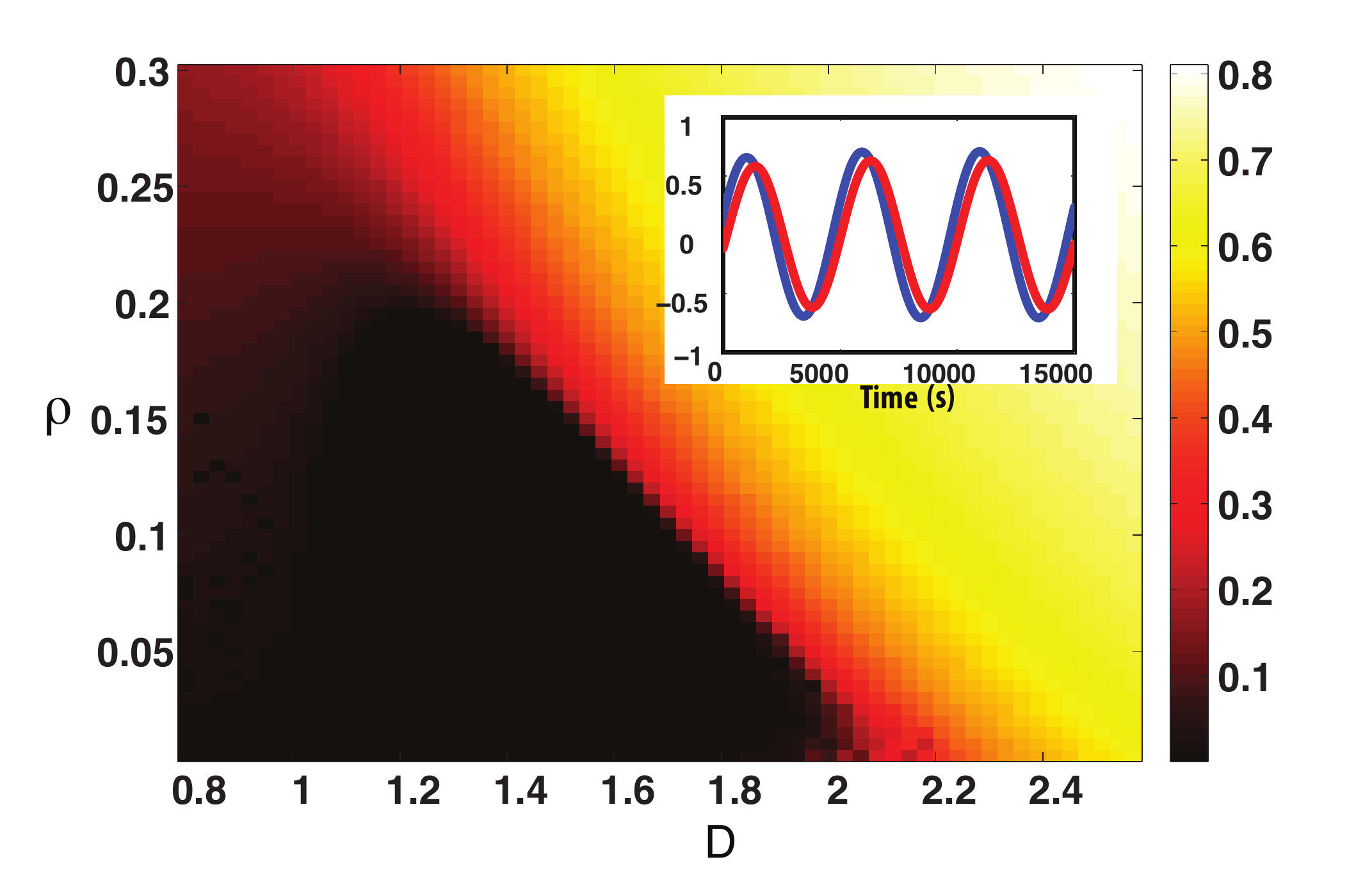}
\caption{(Top) Phase boundaries for homogeneous oscillators in the $\rho$ vs. $D$ plane for various values of $J$, $\omega_0=1$, and $\lambda_0=1$. (Bottom) Heat map of the amplitude of collective oscillations from simulations of $N=40$ identical oscillators, showing the transition from a ``dynamic" death phase with $J=0$ to synchronized oscillations for $D>1$. Inset: Real part of $z$ and $Z$ during low density oscillations showing $\sim 1000$-fold slowing of oscillations (D=2.5, $\rho=0.001$).}
 \label{fig1}
 \end{figure}
 
This equation indicates that there are two qualitative ways to stabilize amplitude death. First, when  $J \gg 1$, the left side is much larger than the right, indicating that oscillations are lost due to degradation of the external medium. Second, even when $J=0$, if the natural frequency of the oscillators is large, $\omega_0 \gg 1$ or the density of oscillators is small, amplitude death can be stable because the right hand side of (\ref{HomoSEQ}) is extremely small. The underlying reason for this is that since $D-\lambda_0 >0$, isolated individual oscillators are silent and synchronization can only occur by transmitting information through the external medium. The external medium has an effective time scale, $(\rho D)^{-1}$, on which it can respond. If $\omega_0$ is large, the medium cannot track the dynamics of the individual oscillators and the death phase is stable. We term this  ``dynamic death'' to indicate that the underlying cause for the loss of oscillations are the dynamical properties of the external medium. Figure \ref{fig1} shows the phase boundaries in this model as a function of $J$, $\rho$, and $D$. Notice the dynamic death region for small $J$ and $ \rho$. We have also confirmed the existence of this phase using numerical simulations (see Figure \ref{fig1}, bottom).  
 
We now analyze (\ref{MainEq}) for the case where the natural frequencies $\omega_j$ are drawn from an even distribution $g(\omega)$ with zero mean. In this case, the system has three phases: an amplitude death phase where all oscillators are quiet; global, synchronized oscillations; and  an incoherent phase where individual elements are oscillating but the oscillations are unsynchronized. For finite densities, we did not numerically observe any unsteady behavior analogous to that seen in  \cite{matthews1991dynamics}. The stability boundary of the amplitude death phase can again be calculated as in the homogenous case by linearizing the system around the death state $z_j=0, Z=0$ (for details see \cite{nextpub}). In the large $N$ limit, the boundaries of stability for the death phase are $D-\lambda_0>0$ and
 \bea
\frac{\rho D + J}{\rho D^2} & = & \int \, d\omega \, g(\omega) \frac{D-\lambda_0}{(D-\lambda_0)^2+ (b-\omega)^2}  \label{BEstab1}\\
\frac{b+\omega_0}{\rho D^2} &=&- \int \, d\omega \, g(\omega) \frac{b-\omega}{(D-\lambda_0)^2+ (b-\omega)^2}.
\label{BEstab2}
\eea
It is useful to consider various limits of these equations. Notice that when $g(\omega)=\delta(\omega)$, these equations reduce to (\ref{HomoEq}) with $a=0$ as expected.  Alternatively, consider the case $\rho \rightarrow \infty$. In this limit, the left hand side of (\ref{BEstab2}) is zero implying that $b=0$, since $g(\omega)$ in an even function. Plugging this into (\ref{BEstab1}) yields a single equation for stability of the death state,
\be
\frac{\rho D + J}{\rho D^2} = \int \, d\omega \, g(\omega) \frac{D-\lambda_0}{(D-\lambda_0)^2+ \omega^2}.
\ee
This equation was derived in \cite{strogatz1991stability, matthews1991dynamics} for the stability boundary  of the death phase in  a system of {\it directly coupled} limit-cycle oscillators. This follows naturally  by noting that in the limit $\rho \rightarrow \infty$,  the external media can respond infinitely quickly. Thus, $Z_{ext}$ is equal to the order parameter of the system, $Z_{ext}=\frac{1}{N} \sum_{j} z_j$, and the model reduces to the one studied in  \cite{strogatz1991stability, matthews1991dynamics, strogatz2000kuramoto}. In this limit, the loss of oscillations is due to the heterogeneity of individual oscillator frequencies and has been termed amplitude death. These two limits show that a single set of equations  (\ref{BEstab1})-(\ref{BEstab2}), capture all four qualitatively distinct types of transitions to synchronized oscillations.

To gain further insight into the system, it is useful to consider the mean-field equations for the system. To do so, we put $z_j=r_je^{i\theta_j}$ and $Z=Re^{i\phi}$ into (\ref{MainEq}) and equate real and imaginary parts:
\bea
\frac{dr_j}{dt} &=& (\lambda_0-D-r_j^2)r_j+ DR\cos{(\phi-\theta_j)} \label{PE1}  \nonumber \\
\frac{d\theta_j}{dt} &=& \omega_j +\frac{DR}{r_j} \sin{(\phi-\theta_j)} \label{PE2} \nonumber \\
\frac{dR}{dt} &=& \frac{\rho D}{N} \sum_{j=1}^N r_j \cos{(\phi-\theta_j)}-(\rho D+J)R \label{PE3} \nonumber \\
\frac{d\phi}{dt} &=& -\omega_0+ \frac{\rho D}{N} \sum_{j=1}^N \frac{r_j}{R} \sin{(\phi-\theta_j)} \label{PE4}.
\eea
We look for uniform rotating solutions whose angular frequency in the lab frame is $\omega_0+b$ by requiring $\frac{dR}{dt}=\frac{dr_j}{dt}=0$ and
$\frac{d\phi}{dt}=\frac{d\theta_j}{dt}=b$ in (\ref{PE4}). In general, these equations cannot be solved analytically. However for the special case $R=0$ corresponding to the death phase one finds that the equations reduce to (\ref{BEstab2}) (see \cite{nextpub}).  This allows us to attach a physical meaning to the parameter $b$. When oscillators are directly coupled to each other ($\rho \rightarrow \infty$), they rotate at the mean frequency $\omega_0$ and $b=0$. In contrast, when oscillators are coupled to each other through the external media, there is an effective ``viscosity '' which slows down the oscillations so that they rotate with an angular frequency $\omega+b$, with $b<0$. Thus, $b$ measures the change in angular frequency of oscillations due to delays induced by the external medium (see Figure \ref{fig1} inset). Furthermore, increasing $J$ decreases the absolute value of $b$ and hence increases the angular frequency. Thus, somewhat surprisingly, the system exhibits positive period-amplitude coupling. This behavior was observed in a population of synthetically engineered bacteria in recent experiments \cite{danino2010synchronized}, though interestingly, the same phenomena occurs as degradation is decreased. The effect of time delays on synchronization of directly coupled oscillators was studied previously  and the equations governing the stability of amplitude death bear some similarity to those found in this work \cite{Ramana1999time, atay2003distributed}. 

When $D<\lambda_0$, the system can be incoherent, where individual oscillators are oscillating in an unsynchronized fashion. The stability equations for the incoherent phase were calculated by generalizing the calculations in \cite{matthews1991dynamics}. We looked for solutions of  (\ref{PE4}) of the form $R=0$, $r_j^2=\lambda_0-D$, and $\theta_j=\omega_jt$. For such solutions, individual oscillators oscillate at their natural frequencies but there are no coherent oscillations. We calculated the stability boundary for incoherence by checking the stability of the state to small perturbations (see \cite{nextpub}). We find that for distributions $g(\omega)$, there is a tri-critical point on the line $D=\lambda_0$ where the incoherent phase, the synchronized oscillation phase, and the death phase meet.

\begin{figure}[t]
 \includegraphics[width=8cm]{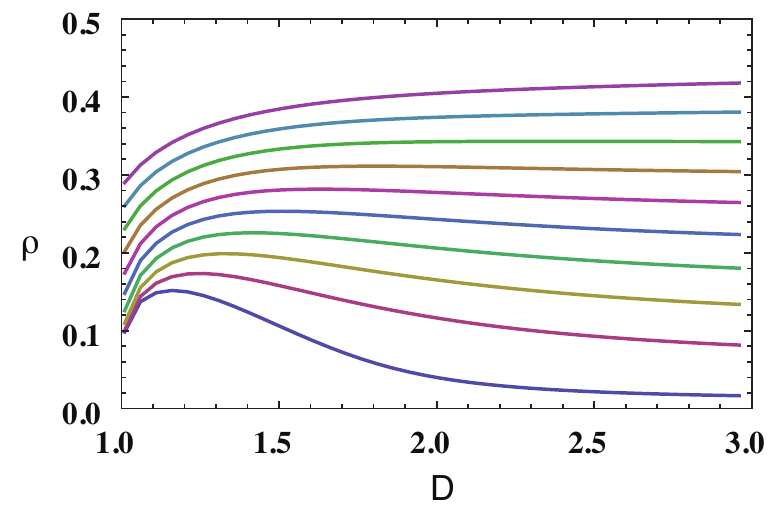}
 \includegraphics[width=8cm]{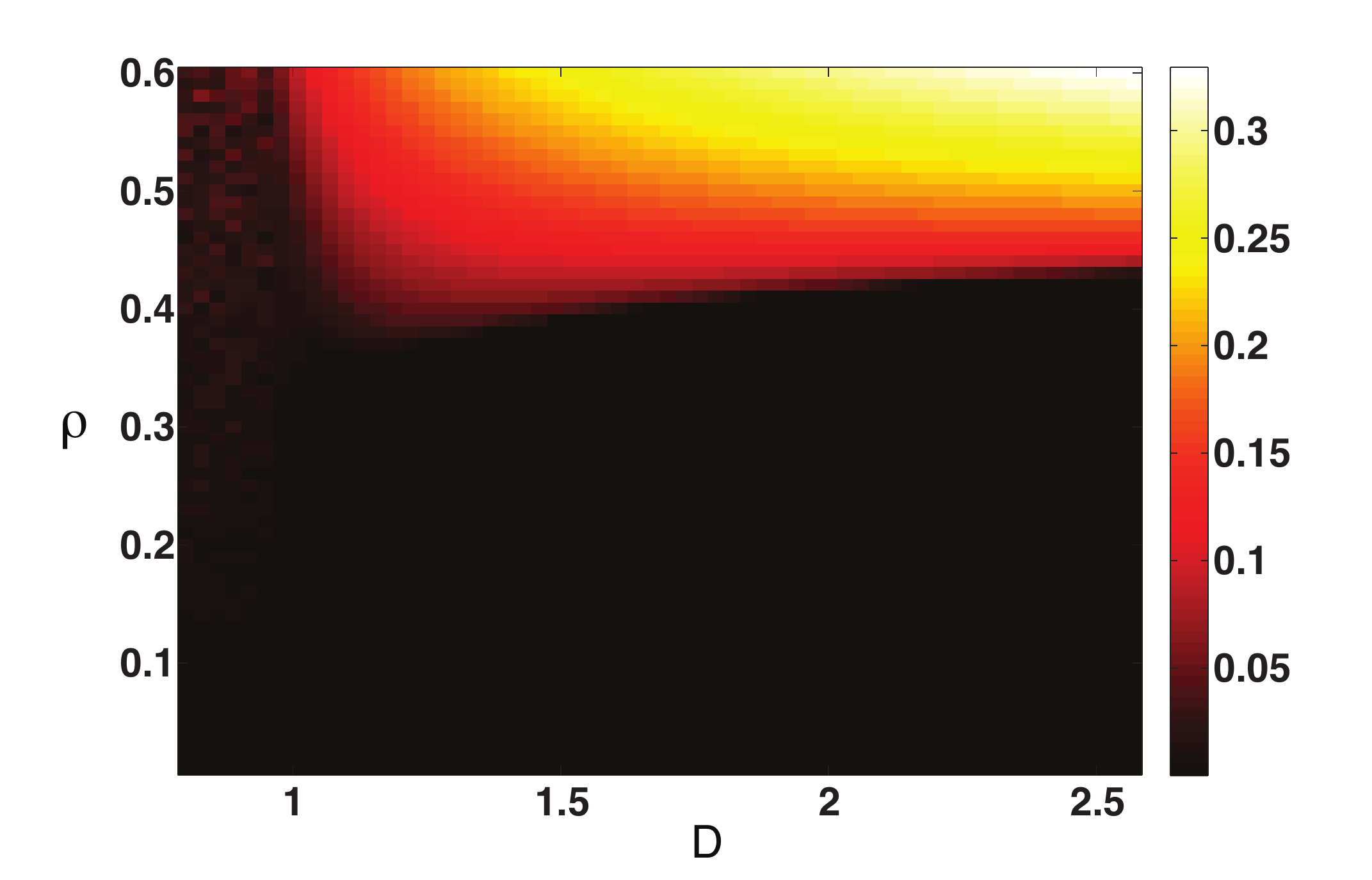}
\caption{(Top) Phase boundaries in $\rho$ vs $D$ plane for heterogeneous oscillators drawn from a rectangular distribution ( $\Gamma=0.6$, $\omega_0=1$, and $\lambda=1$ and $J= .05-0.5$ from bottom to top). (Bottom) Heat map of the amplitude of collective oscillation from simulations of $N=100$ oscillators drawn from a rectangular frequency distribution with parameters above for $J=0.5$. Notice the transitions from an incoherent phase with $D<1$ to synchronized oscillations or death. Incoherence phase  (spotted red region for $D<1$) can be identified by the  $1/N^{1/2}$ fluctuations of the order parameter.}
 \label{fig2}
 \end{figure}

The derivation presented above is for arbitrary $g(\omega)$. When $g(\omega)$ is either a rectangular or Lorentzian distribution, we can perform the integrations in (\ref{BEstab2}) explicitly (see \cite{nextpub}). The answers are particularly simple for a Lorentzian distribution, $ g(\omega)= \frac{1}{\pi} \frac{\Gamma}{\Gamma^2 +\omega^2}$, and are given by
\bea
\frac{\rho D+J}{\rho D^2} &=& \frac{D-\lambda_0+ \Gamma}{(D-\lambda_0+ \Gamma)^2+b^2} \nonumber\\
\frac{b+i\omega_0}{\rho D^2} &=& -\frac{b}{(D-\lambda_0+ \Gamma)^2+b^2}.
\label{BEstabLorentz}
\eea
These equations are identical to the homogenous case, (\ref{HomoEq}), except with $\lambda_0 \rightarrow \lambda_0-\Gamma$ on the right hand side. Thus, heterogeneity  ``renormalizes" the distance individual units are from their Hopf bifurcation.  An analogous set of equations--albeit more unwieldy--can also be derived for a rectangular frequency distribution, and Figure \ref{fig2} shows the phase boundaries for this case as a function of $\rho$, $D$, and $J$. As expected, for $D>\lambda_0$, the death phase and synchronized oscillations are both possible. For large $D$, as density is increased across the transition, the amplitude of the synchronized oscillations rises sharply with density. For smaller $D$, this rise in amplitude is less pronounced. When $D < \lambda_0$, one also sees a Kuramoto-like transition from incoherent to synchronized oscillations. The same qualitative behavior was observed in recent experiments on BZ catalytic particles with a distribution of natural  frequencies  \cite{taylor2009dynamical,  tinsley2010dynamical}.

In this letter, we considered the physics of limit-cycle oscillators diffusively coupled through an external media. We have found that there are four qualitatively different types of density-dependent transitions to synchronized oscillations including a new type of transition we term ``dynamic death", where the dynamics of the external medium are too slow to support oscillations. This simple model captures many qualitative features seen in a variety of experiments on oscillators coupled diffusively through an external medium. For example, it was previously argued that when all oscillators are identical, the model is a good description of glycolitic oscillations in suspensions of yeast cells. The model also shows how large amplitude oscillations can emerge as one varies the density and how this behavior crosses-over into a  Kuramoto-like transition as $D$ is decreased (see Fig. \ref{fig2}). These qualitative features are in good agreement with  \cite{taylor2009dynamical,  tinsley2010dynamical}. Finally, the model also captures many of the mean-field properties of coupled synthetically-engineered bacteria, including the sudden emergence of oscillations and scaling of amplitude and period of oscillations as one changes the external degradation rate $J$. However, unlike  \cite{danino2010synchronized}, the period and amplitude of the oscillations decrease with increasing $J$. This discrepancy likely arises from the highly non-linear nature of the ``degrade-and-fire'' oscillations \cite{mather2009delay} characterizing the synthetic bacteria. Despite this discrepancy, our results  suggest that properly constructed simple models may be able to capture interesting, qualitative behaviors of coupled oscillators. In the future, it will be interesting to directly relate this simple model to more detailed models \cite{mather2009delay} and extend the simple mean-field model of dynamical quorum sensing explored here to include spatial effects.

\begin{acknowledgments}
{\bf Acknowledgments} We thank Troy Mestler and Thomas Gregor for useful discussions. This work was partially supported by NIH Grants K25GM086909 (to P.M.). DS was partially supported by DARPA grant HR0011-05-1-0057 and NSF grant PHY-0957573. 

\end{acknowledgments}

\bibliography{refsmain}

\section{Supporting Information: Equations for stability of amplitude death}

In this section, we derive equations for the stability of the amplitude death phase. We start with the equations
\bea
\frac{dz_j}{dt}&=& (\lambda_0+i \omega_j - |z_j|^2)z_j -D(z_j-Z)   \label{aOE1} \\
\frac{dZ}{dt} &=& \frac{\rho D}{N} \sum_j( z_j -Z) -(J+ i \omega_0) Z \label{aOE2}.
\eea
where we have transformed to a frame rotating at the mean frequency, $\omega_0$.
To calculate the stability of amplitude death, we linearize these equations around the solution $z_j=Z=0$. Doing so yields the equations,
\bea
\frac{d \delta z_j}{dt}&=& (\lambda_0+ i\omega_j -D) \delta z_j-D \delta Z  \\
\frac{d \delta Z}{dt} &=& \sum_j \frac{\rho D}{N} \delta z_j - (\rho D + J + i \omega_0)\delta Z .
\label{lOE} 
\eea
These equations can be written in matrix form as
\be
\left(\begin{array}{c} \dot{\delta z_1}\\
\dot{\delta z_2} \\
\cdot\\
\cdot \\
\dot{\delta z_N} \\
\dot{\delta Z}
\end{array} \right) = 
\left(\begin{array}{c c c c c c}
(\lambda_0 -D +i \omega_1) & 0 &\cdot & \cdot & 0 & -D \\
 0  & (\lambda_0 -D +i \omega_2)  &\cdot & \cdot & 0 & -D \\
\cdot & \cdot & \,\, \cdot &\,\, \cdot &\cdot & \cdot \\
\cdot & \cdot & \cdot & \cdot &\cdot & \cdot \\
 0  &  0 & \cdot & \cdot &(\lambda_0 -D +i \omega_N)  & -D \\
\frac{\rho D}{N} & \frac{\rho D}{N} & \cdot & \cdot & \frac{\rho D}{N} & -(\rho D +J+ i \omega_0)
 \end{array} \right) 
\left(\begin{array}{c} \delta z_1\\
\delta z_2 \\
\cdot\\
\cdot \\
\delta z_N \\
\delta Z
\end{array} \right)
\label{LME}
\ee
Denote the matrix on the right hand side of (\ref{LME})  by $M$ for notational convenience.
Stability requires the eigenvalues,  $\mu$, of  $M$  to satisfy $Re[ \mu] <0$. First notice that stability requires $Re[Tr(M)]<0$. This gives the condition
\be
(\lambda_0-D)-\frac{(\rho D+J)}{N} > 0.
\label{DET}
\ee

We can also calculate the eigenvalues using the  characteristic equation of the matrix, $Det(\mu I-M)=0$. A straightforward calculation yields
\be
(\mu+(\rho D+ J +i \omega_0)) \prod_{j=1}^N (\mu-(\lambda_0-D +i \omega_j) -\frac{\rho D^2}{N}\sum_{s=1}^N \prod_{j=1, j \neq s}^N (\mu-(\lambda_0-D +i \omega_j) =0
\ee
In order to take the thermodynamic limit, we rewrite this equation as
\be
(\mu+(\rho D+ J +i \omega_0)) = \frac{\rho D^2}{N} \sum_{s=1}^N \frac{1}{\mu-(\lambda_0-D +i \omega_j)}
\label{DEE}
\ee
In the thermodynamic limit $N \rightarrow \infty$ but with $\rho$ held fixed, (\ref{DET}) and (\ref{DEE}) become, respectively,
\be
(\lambda_0-D)<0
\ee
and
\be
\mu+(\rho D+ J +i \omega_0)) =  \rho D^2 \int d\omega \, \frac{g(\omega)}{\mu-(\lambda_0-D +i \omega)},
\ee
where we have replaced the sum by an integral over the distribution function $g(\omega)$ for the oscillator frequencies.  In practice, it is often helpful to write this as two real equations. Substituting 
$\mu=a+ib$ yields the coupled equations. 
\bea
\frac{a+\rho D + J}{\rho D^2} &=& \int \, d\omega \, g(\omega) \frac{ a+D-\lambda_0}{(a+D-\lambda_0)^2+ (b-\omega)^2} \\
\frac{b+\omega_0}{\rho D^2} &=&- \int \, d\omega \, g(\omega) \frac{b-\omega}{(a+D-\lambda_0)^2+ (b-\omega)^2}
\label{complexEquations}
\eea
Stability requires that all solutions of these equations obey $a \leq 0$. By considering the mean-field equations derived below (and considering various limiting cases of the equations above), it is clear that the stability boundary can be found by putting $a=0$ in the above equations, i.e. there exists at most one solution with positive real part. 

We illustrate this explicitly for the case of uniform frequencies, where all the oscillators are identical $g(\omega)=\delta(\omega)$  (since by assumption the center of the distribution is around $\omega=0$). Then, the equations above can be rewritten as
 \begin{eqnarray}
a+B=\frac{\rho D^2(a+A)}{(a+A)^2+b^2} \label{HomoEq} \\
b+\omega_0=\frac{-\rho D^2 b}{(a+A)^2+b^2}.
\label{BEstabdelta}
\end{eqnarray}
where $A=D-\lambda_0$ and $B=D\rho+J$. Dividing the equations and solving for $b$ in terms of $a$ to get $b(a)=-\frac{\omega_0(a+A)}{2a+B+A}$ results in an equation for $a$ that can be analyzed graphically. Plotting the left-hand and right-hand sides of equation (\ref{BEstabdelta}) in Figure \ref{fig1}, we see that there always exists a solution with negative real part. Recall that the characteristic equation is quadratic, assuring at most two solutions. The second solution can be either positive or negative and, since there are no other solutions with positive real part, setting $a=0$ safely identifies the phase boundary.

\begin{figure}[t]
\includegraphics[width=10cm]{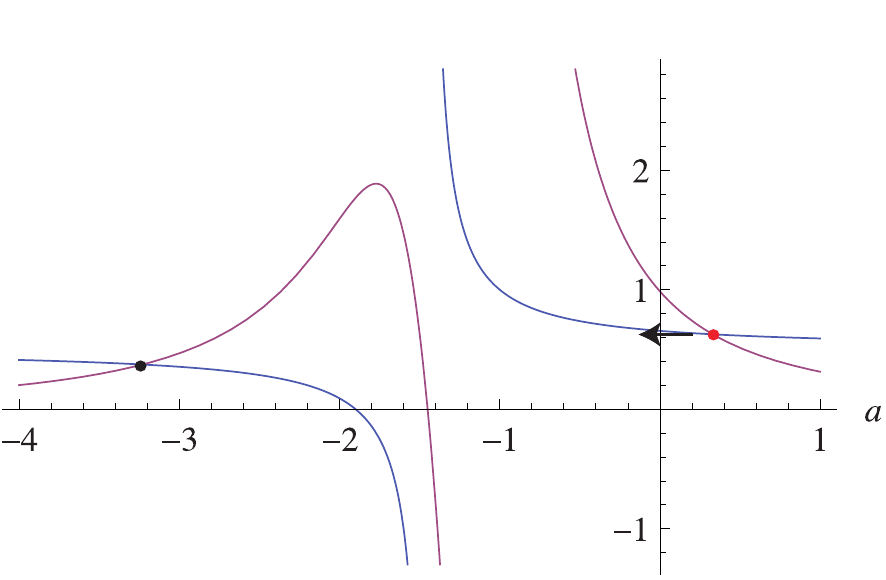}
\caption{Graphical analysis of the structure of the solutions to eq. (\ref{HomoEq}) and (\ref{BEstabdelta}). There always exists a solution with $a<0$, shown as a black dot. A second solution, shown as a red dot, can have positive or negative $a$, and thus solving for $a=0$ properly identifies the phase boundary.}
\label{fig1}
\end{figure}
 
Putting $a=0$ for the case with heterogeneous frequencies results in a pair of coupled integral equations that determine the boundary of stability of the death phase:
 \bea
\frac{\rho D + J}{\rho D^2} & = & \int \, d\omega \, g(\omega) \frac{D-\lambda_0}{(D-\lambda_0)^2+ (b-\omega)^2} \label{BEstab2}\\
\frac{b+\omega_0}{\rho D^2} &=&- \int \, d\omega \, g(\omega) \frac{b-\omega}{(D-\lambda_0)^2+ (b-\omega)^2}
\label{BEstab}
\eea

We can consider various limiting cases of the equations above. For uniform frequencies, the resulting equations are precisely those derived by Del Monte and co-workers. 
We can also consider the $\rho \rightarrow \infty$ limit. This corresponds to the case where the oscillators are {\it directly} coupled through the mean-field parameter. In this case, the left-hand side of the second equation in (\ref{BEstab}) is zero. Since $g(\omega)$ is an even function, this implies $b=0$. Plugging this into the top equation in (\ref{BEstab}) and taking the limit $\rho \rightarrow \infty$ yields the equation
\be
\frac{1}{D} = \int d\omega \, g(\omega) \frac{D-\lambda_0}{(D-\lambda_0)^2+ \omega^2}.
\ee
This is precisely the equation derived  by Mirollo {\it et al.} for direct all-to-all coupling.

\section{Mean-field equations for frequency locking}
Here we derive the mean-field equations for the model and show that they reproduce the results from the determinant calculations. We begin again with the equations 
\bea
\frac{dz_j}{dt}&=& (\lambda_0+i \omega - |z_j|^2)z_j -D(z_j-Z)   \label{OE1} \\
\frac{dZ}{dt} &=& \frac{\rho D}{N} \sum_j( z_j -Z) -(J+ i \omega_0) Z \label{OE2}.
\eea
In writing these equations we have assumed that $\omega_j$ are drawn from a normalized, even, probability distribution $g(\omega_j)$. Now define $z_j=r_je^{i\theta_j}$ and $Z=Re^{i\phi}$. Then, we can rewrite these equations  in polar coordinates to get 
\bea
\frac{dr_j}{dt} &=& (\lambda_0-D-r_j^2)r_j+ DR\cos{(\phi-\theta_j)} \label{PE1} \\
\frac{d\theta_j}{dt} &=& \omega_j +\frac{DR}{r_j} \sin{(\phi-\theta_j)} \label{PE2}\\
\frac{dR}{dt} &=& \frac{\rho D}{N} \sum_{j=1}^N r_j \cos{(\phi-\theta_j)}-(\rho D+J)R \label{PE3} \\
\frac{d\phi}{dt} &=& -\omega_0+ \frac{\rho D}{N} \sum_{j=1}^N \frac{r_j}{R} \sin{(\phi-\theta_j)} \label{PE4}
\eea

We now look for uniform, rotating, locked solutions where $\frac{dR}{dt}=\frac{dr_j}{dt}=0$ and
$\frac{d\phi}{dt}=\frac{d\theta_j}{dt}=b$. In this case, the position of each oscillator is determined purely by its frequency, so we can regard each oscillator as a function of $\omega$. Using equations (\ref{PE1}) and (\ref{PE2}) and plugging in the expressions for the derivatives on the left hand side  one can easily show that 
\be 
((\omega-b) \cot{(\theta-\phi)} +\lambda_0-D) (1+\cot^2{(\theta-\phi)})(\omega-b)^2=D^2R^2 
\label{SCE}
\ee
Now, in the thermodynamic limit plugging in the desired solutions yields
\bea
R(\rho D +J)= \rho D\int d\omega g(\omega) r(\omega) \cos{(\phi-\theta(\omega))} \label{SC1}\\
b+\omega_0= \rho D\int d\omega g(\omega)\frac{ r(\omega)}{R}\sin{(\phi-\theta(\omega))} \label{SC2},
\eea
where we have written $r(\omega)$ and $\theta(\omega)$ to emphasize the amplitude and phase of each oscillator is a function of only the frequency. Using equation (\ref{PE2}) and the fact that $\frac{d\theta_j}{dt}=b$ yields
\be
r = \frac{DR \sin{(\theta(\omega)-\phi)}}{(\omega-b)}.
\ee
Plugging this into (\ref{SC1}) and (\ref{SC2}) gives the equations
\bea
\frac{(\rho D+ J)}{\rho D^2} &=&\int d\omega g(\omega) \frac{  \sin{(\theta(\omega)-\phi)} \cos{(\theta(\omega)-\phi)}}{\omega-b} \label{MF1} \\
\frac{b+\omega_0}{\rho D^2} &=&- \int d\omega g(\omega) \frac{  \sin{(\theta(\omega)-\phi)} \sin{(\theta(\omega)-\phi)}}{\omega-b} \label{MF2}
\eea
Combined (\ref{MF1}), (\ref{MF2}), and (\ref{SCE}) define the mean-field equations for the system for frequency locking with amplitude $R$.

In general, we cannot solve this analytically because (\ref{SCE}) is a cubic equation in $\cot$. However, for the special case $R=0$ (oscillator death), we have the unique solution to (\ref{SCE}) that 
\be
\tan{(\theta(\omega)-\phi)} =\frac{\omega-b}{\lambda_0-D}.
\ee
Plugging this into the equations (\ref{SC1}) and (\ref{SC2}) yields the equations for amplitude death to be stable 
\bea
\frac{(\rho D+ J)}{\rho D^2} &=&\int d\omega g(\omega) \frac{D-\lambda_0}{(D-\lambda_0)^2+ (b-\omega)^2} \label{AD1}  \label{SE1}\\
\frac{(b+\omega_0)}{\rho D^2} &=& -\int d\omega g(\omega) \frac{(b-\omega)}{(D-\lambda_0)^2+ (b-\omega)^2} \label{SE2}
\eea
These are precisely our equations obtained from the derivative expansion. It gives as an interpretation of the imaginary part of the eigenvalue $b$ as the shift of the frequency of the collective oscillations from $\omega_0$. This should be contrasted with the case where oscillators are directly coupled and $b=0$.

Finally, we can ask intersect the stability boundary $D=\lambda_0$. We can do this by taking the limit $(D-\lambda) \rightarrow 0$ in the equations above. A straightforward calculation shows that the equations reduce to 
\bea
\frac{(\rho D+ J)}{\rho D^2} &=& \pi g(b) \nonumber \\
\frac{(b+\omega_0)}{\rho D^2} &=& P \left[ \int_{-\infty}^\infty d\omega \, \frac{g(\omega)}{\omega-b}\right],
\label{ADDeqlam}
\eea
where the $P$ denotes the principal value.

\section{Mean Field Equations for incoherent state}

In this section, we derive the mean field equations for the stability of the incoherent state when $D < \lambda$. In this case, we look for solutions of  (\ref{PE1})-(\ref{PE4}) of the form $R=0$, $r_j^2=\lambda_0-D$. In such a solution, individual oscillators oscillate at their natural frequencies but there is no coherent oscillations.

We now examine the stability of the coherent state. To do so, we follow the analysis developed by Mathews et al. Define a density function $\rho(r, \theta, \omega, t)$ so that the fraction of oscillators of frequency $\omega$ between $r$ and $r+dr$ and between $\theta$ and $\theta+d\theta$ is $\rho r d\theta dr$. The evoloution for $\rho$ is given by the continuity equation
\be
\frac{\partial \rho}{\partial t} + \vec{\nabla} \cdot (\rho \vec{\nu}) =0
\ee
where $\vec{\nu}$ is the velocity of oscillators given by $\vec{\nu}=(\dot{r}, r \dot{\theta})$. Substituting (\ref{PE1}) and (\ref{PE2}) gives
\be
\frac{\partial \rho}{dt} + \frac{1}{r} \frac{\partial}{\partial r}\left( \rho \left[r^2(a^2-r^2)+KRr\cos(\theta-\phi)\right]\right)+\frac{1}{r} \frac{\partial}{\partial \theta}\left(\rho\left[r\omega-KR\sin(\theta-\phi)\right]\right)=0,
\ee
where $a^2=(\lambda_0-D)$. In the incoherent state
\be
\rho= \frac{\delta(r-a)}{2\pi r}.
\ee

We now consider a small perturbation in the radial and angular directions and check when the density is stable to these perturbations. In particular, consider
\be
\rho= \delta(r-a -\epsilon r_1(\theta, \omega, t))\left(\frac{1}{2\pi r} + \epsilon f_1 (\theta, \omega, t)\right).
\label{rhofull}
\ee
For such a perturbation, by the chain rule we have
\be
\dot{r} = \epsilon \frac{\partial r_1}{\partial \theta} \cdot{\theta} + \epsilon \frac{\partial r_1}{\partial t}.
\ee
Writing $R=\epsilon R_1$, substituting in (\ref{PE1}) and (\ref{PE2}), and keeping terms first order in $\epsilon$ yields
\be
-2 a^2 r_1+DR_1\cos(\theta-\phi) =\omega \frac{\partial r_1}{\partial \theta} +  \frac{\partial r_1}{\partial t}.
\ee
We seek solutions in which $R_1$ and $r_1$ are proportional to $e^{(\lambda +ib)t}$ and we find that $r_1$ must obey the equation
\be
\omega \frac{\partial r_1}{\partial \theta} +(\lambda+ib+2a^2)r_1 =DR_1 \cos(\theta-\phi).
\ee
The solution for $r_1$ which is periodic in $\theta$ is of the form
\be
r_1=A \cos(\theta-\phi)+B \sin(\theta-\phi),
\label{r1sol}
\ee
where
\bea
A &=& \frac{DR_1(\lambda+i b+ a^2)}{\omega^2+(\lambda+ib+2a^2)^2} \\
B &=&  \frac{DR_1 \omega }{\omega^2+(\lambda+ib+2a^2)^2}
\eea
We now consider the small angular perturbations. We can substitute $\rho=\delta(r-a)[1/2\pi r + \epsilon f_1]$ into the continuity equation and keep terms linear in $\epsilon$ to get
\be
\frac{\partial f_1}{\partial t} + \omega \frac{\partial f_1}{\partial \theta} - \frac{KR_1 \cos (\theta -\phi)}{2\pi a^2}=0
\ee
Assuming the periodic solution is proportional to $e^{(\lambda +ib)t}$ as above, one finds
\be
f_1= C \cos(\theta-\phi)+ D \sin(\theta-\phi)
\label{f1sol}
\ee
with
\bea
C &=& \frac{DR_1(\lambda+i b)}{2 \pi a^2(\omega^2+(\lambda+ib)^2)} \\
D &=&  \frac{DR_1 \omega }{2 \pi a^2 (\omega^2+(\lambda+ib)^2)}
\eea

We can now rewrite the steady-state equations stemming from  (\ref{PE3}) and (\ref{PE4}) in terms of the density to get 
\bea
\frac{\rho D+ J}{\rho D} R &=& \int_{-\infty}^{\infty}\int_0^\infty \int_0^{2\pi} r \cos{(\theta-\phi)} \rho \, r \, d\theta \, dr \, g(\omega)\, d\omega \nonumber \\
\frac{b+\omega}{\rho D} &=& \int_{-\infty}^{\infty}\int_0^\infty \int_0^{2\pi} r \sin{(\theta-\phi)} \rho \, r \, d\theta \, dr \, g(\omega)\, d\omega,
\eea
where we have used that the order parameter for the solutions is chosen so that $\frac{d \phi}{dt}=b$.
Plugging in (\ref{rhofull}), (\ref{r1sol}), and (\ref{f1sol}), and keeping terms first order in $\epsilon$,
\bea
\frac{2(\rho D+ J)}{\rho D} &=& \int_{-\infty}^\infty \frac{\lambda+ib}{(\lambda+ib)^2 + \omega^2}\,g(\omega) d \omega + \int_{-\infty}^\infty \frac{\lambda+ib+2a^2}{(\lambda+ib+2a^2)^2 + \omega^2}
\,g(\omega) d \omega \\
\frac{2(b+\omega_0)}{\rho D^2}&=& \int_{-\infty}^\infty \frac{\omega}{(\lambda+ib)^2 + \omega^2}\,g(\omega) d \omega + \int_{-\infty}^\infty \frac{\omega}{(\lambda+ib+2a^2)^2 + \omega^2}
\,g(\omega) d \omega
\eea
The bifurcation condition requires that $\lambda=0$. So the stability boundary is given by setting $\lambda=0$ in the equation above. This gives (using usual relationships for principal values of integrals  in the limit $\lambda =0^+$)
\bea
\frac{2(\rho D+ J)}{\rho D} &=& \pi g(b)
+ \int_{-\infty}^\infty \frac{ib+2a^2}{(0^+ + ib+2a^2)^2 + \omega^2}
\,g(\omega) d \omega \\
\frac{2(b+\omega_0)}{\rho D^2}&=&P \left[ \int_{-\infty}^\infty d\omega \, \frac{g(\omega)}{\omega-b}\right] + \int_{-\infty}^\infty \frac{\omega}{(0^+ +ib+2a^2)^2 + \omega^2}
\,g(\omega) d \omega
\eea
where $P$ denotes the principal value. For the line $D=\lambda_0$ (i.e. $a=0^+$) these equations reduce to (\ref{ADDeqlam}) showing that the incoherence joins the corner of the death state.

\section{Expressions for Lorentzian \& Rectangular Distributions}
\subsection{Lorentzian Distributions}

We now derive analytic expression for the boundary for the special case where
 $g(\omega)$ is a Lorentzian,
\be
g(\omega)= \frac{1}{\pi} \frac{\Gamma}{\Gamma^2 +\omega^2}
\label{gdef}
\ee
For a Lorentzian distribution, the equations are particularly simple because the Fourier transform is a simple exponential:
\be
\hat{g}(p)= e^{-|p|\Gamma}.
\ee
We now plug this into the equations for the stability of amplitude death (\ref{BEstab}) and use the fact that these equations can be thought of as a convolution for $b$.  A straightforward calculation then shows that the resulting equations for the stability boundary are identical to the case where $g(\omega)=\delta(\omega)$, given by (\ref{BEstabdelta}), except with $D-\lambda_0 \rightarrow D-\lambda_0 + \Gamma$,
\bea
\frac{\rho D+J}{\rho D^2} &-& \frac{D-\lambda_0+ \Gamma}{(D-\lambda_0+ \Gamma)^2+b^2} \nonumber\\
\frac{b+i\omega_0}{\rho D^2} &=& -\frac{b}{(D-\lambda_0+ \Gamma)^2+b^2}.
\label{BEstabLorentz}
\eea
Thus $\Gamma$ has the intriguing effect of decreasing the effective $\lambda_0$, thereby pulling the individual oscillators closer to their supercritical Hopf bifurcation.
\subsection{Rectangular Distributions}

The integrals can also be performed for the case where $g(\omega)$ is drawn from a rectangular distribution,
\bea
g(\omega) &= \, 1/\Gamma  &\text{\hspace{0.5in} if } -\Gamma/2<\omega<\Gamma/2 \nonumber\\
&= \,0 \,&\text{\hspace{0.5in} otherwise} 
\eea
In this case, the integrals in (\ref{BEstab2}) and (\ref{BEstab}) can be performed and yield the equations
\begin{align}
\frac{a+A}{\rho D^2}= \frac{1}{2\Gamma}&\left( \arctan[ (b+\Gamma)/(a+B)] -\arctan[ (b-\Gamma)/(a+B)] \right) \\
&\frac{b+\omega_0}{\rho D^2}= \frac{1}{2\Gamma}\log\left[ \frac{(b+\Gamma)^2+(a+B)^2}{(b-\Gamma)^2+(a+B)^2}\right]
\end{align}

\end{document}